\documentclass[pre,amsmath,showpacs]{revtex4-1}
\newcommand{\be}{\begin{equation}}
\newcommand{\ee}{\end{equation}}
\newcommand{\bea}{\begin{eqnarray}}
\newcommand{\eea}{\end{eqnarray}}

\usepackage{graphicx} 
\usepackage{color}
\usepackage{bm}
\usepackage{epsfig}
\usepackage{latexsym}
\usepackage{pifont}
\usepackage{nameref,hyperref}

\begin{document}

\title{Run-and-tumble motion with step-like responses to a stochastic input}
\author{ Subrata Dev and Sakuntala Chatterjee}
\affiliation{Department of Theoretical Sciences, S. N. Bose National Centre for Basic Sciences, Block  JD, Sector  III, Salt Lake, Kolkata  700106, India. }

\begin{abstract}
We study a simple run-and-tumble random walk whose switching frequency from run mode to tumble mode and the reverse depend on a stochastic signal. We consider a particularly sharp, step-like dependence, where the run to tumble switching probability jumps from zero to one as the signal crosses a particular value (say $y_1$) from below. Similarly, tumble to run switching probability also shows a jump like this as the signal crosses another value ($y_2 < y_1$) from above. We are interested in characterizing the effect of signaling noise on the long time behavior of the random walker. We consider two different time-evolutions of the stochastic signal. In one case, the signal dynamics is an independent stochastic process and does not depend on the run-and-tumble motion. In this case we can analytically calculate the mean value and the complete distribution function of the run duration and tumble duration.  In the second case, we assume that the signal dynamics is influenced by the spatial location of the random walker. For this system, we numerically measure the steady state position distribution of the random walker. We discuss some similarities and differences between our system and {\it Escherichia coli} chemotaxis, which is another well-known run-and-tumble motion encountered in nature.   
\end{abstract}
\maketitle

\section{Introduction}
\label{sec:intro}

Run-and-tumble motility is widely used by a large variety of microorganisms.  Prokaryotic cells like {\it Escherichia coli}, {\it Salmonella typhimurium}, {\it Bacillus subtilis}, {\it Rhodobacter sphaeroides} and {\it Serratia marcescens} navigate in their environment by alternatively switching between a run mode and a tumble mode \cite{salmonela, bacilus, rhodo, serra}. Even eukaryotic organisms like {\it Chlamydomonas rheinhartii} or {\it Tritrichomonas foetus} are known to use run-and-tumble strategy to move around \cite{chlamy, tritrich, berg2003_flagella}. Out of all these cells, the motion of {\it E. coli} is the most well-characterized one \cite{bergbook,celani2010,celani2011}. During the run mode, when {\it E. coli} cell moves in one direction with a fixed speed, the flagellar motors in the cell rotate in the counter-clockwise (CCW) direction which helps the formation of a flagellar bundle and propels the cell forward. When some of the motors start rotating in the clockwise (CW) direction, the corresponding flagella come out of the bundle and the bundle gets dispersed, which results in tumbling of the cell\cite{adler1973,berg3d}. During a tumble mode, the cell does not have significant displacement, but this mode helps the cell to reorient itself and choose a new direction for the next run.

The rotational bias of the flagellar motors is controlled by phosphorylated motor protein CheY-P inside an {\it E. coli} cell, which binds to the motors and increases their CW bias. Importantly, the dependence of CW bias on CheY-P concentration\cite{bren} is very sensitive and experiments measure an almost sigmoidal dependence\cite{cluzel2000}, where CW bias changes sharply from $0$ to $1$ as CheY-P concentration varies within a small range. Since CW bias is the direct measure of tumbling rate, this means the probability for a cell to tumble is vanishingly small when CheY-P level falls below a certain value, and when CheY-P level goes slightly higher, the tumbling probability becomes very close to $1$ and the cell almost always tumbles.

\begin{figure}[h!]
\includegraphics[scale=1.2]{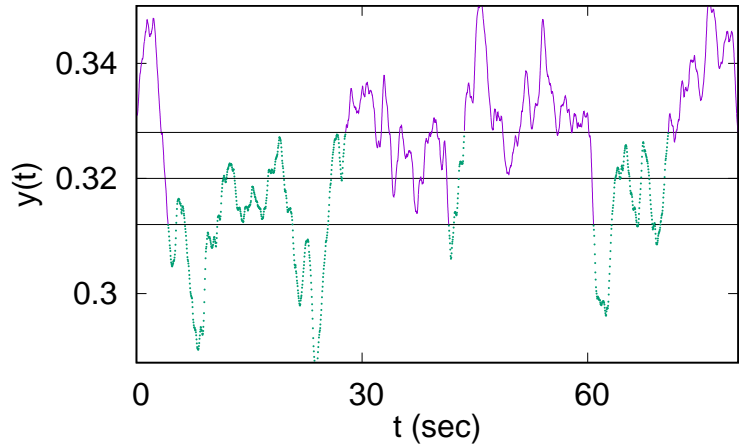}
\caption{A typical time series of the signal $y(t)$. The purple (continuous line) segments correspond to tumbles and the green (dashed line) segments correspond to runs. The values $y_0$ and $y_0 \pm \Delta/2$ are shown by the three horizontal lines. We have used  $y_0=0.32$ and $\Delta = 0.016$ here. Every time $y(t)$ exits the range through a boundary different from the one it had used to enter the range, a switch happens. }
\label{fig:model}
\end{figure}

These observations give rise to a more general and interesting theoretical question: what is the effect of a sharp or sigmoidal switching response on a simple run-and-tumble motion?  This question cannot be addressed within the widely used coarse-grained description of run-and-tumble motion where the system is studied over a time-scale which is much longer than a typical run duration \cite{cates}. This approach is useful in describing the motion in terms of an effective drift velocity and diffusion constant in the long time regime when a large number of tumbling events have already taken place \cite{cates,sc,schnitzer}. However, to understand the effect of a sharp switching response between the run mode and tumble mode, one needs a more microscopic model of a run-and-tumble dynamics and in this work we have developed and studied such a model. We consider a simple run-and-tumble random walker whose switching probabilities between run and tumble modes depend on a certain (stochastic) input signal. To study the system in the simplest possible setting, we consider only two values of the switching probabilities, $0$ and $1$. An infinitely sharp response curve would mean that as the input signal level crosses a certain threshold value, the switching probability jumps from $0$ to $1$. However, such a sharp response means that within a finite time-interval there can be an infinite number of switching events which is unphysical. So we introduce a small range of width $\Delta$ around the threshold value, such that the probability to switch from run to tumble mode is zero (one) as the input signal stays below (above) this range. In other words, run to tumble switch happens, as the input signal crosses the $\Delta$ range from below and goes above it. Once the random walker is in the tumble mode, the tumble to run switch happens with probability one when the input signal decreases and falls below the $\Delta$ range. Thus the two switches happen at two different values of the input signal level, which are separated by the range $\Delta$. When the input signal has any other value, no switching event takes place and the random walker just continues in its current mode. In Fig. \ref{fig:model} we present a typical example.

{\color{blue}  }

We are interested to characterize the motion of the random walker in the long time limit, and to understand how the fluctuations present in the input signal affect the motion. We consider two types of cases here: one in which the dynamics of the input signal is an independent process, and another in which the time-evolution of the signal is also influenced by the position of the random walker. Since our study is motivated from the run-and-tumble motion found in several organisms in nature, including {\it E. coli}, we choose the time-evolution of the signal from the well-studied physical system of chemotactic pathway of an {\it E. coli} cell. The CheY-P level inside the cell fluctuates with time and we consider this to be our input signal. In presence of a concentration gradient of the nutrient, the CheY-P dynamics depends on the local nutrient concentration, and hence on the cell position. However, when the cell moves in a homogeneous nutrient environment, CheY-P fluctuation does not involve the cell position. In the latter case, various quantities can be calculated exactly. Using the fact that the switching events can have only probabilities $0$ and $1$, we show that it is possible to  describe the switching as a first passage process. From this, the probability to observe a certain run (or tumble) duration can be calculated exactly. We also calculate average run and tumble duration and show that both decrease as a function of the signaling noise strength. Our Monte Carlo simulations agree well with our analytical calculations. In the case when the signal dynamics also depends on the position of the random walker, we find the steady state distribution of the random walker position, for a given nutrient concentration profile in the medium, and show that it is more likely to find the random walker in a region where the nutrient concentration is higher. This shows that even within this very simple version of run-and-tumble, where switching probabilities between the two modes are either $0$ or $1$, the basic signature of chemotaxis, which is to find the walker in regions with more food with more likelihood, is recovered.

This paper is organized as follows. In Sec. \ref{sec:model6} we study run-and-tumble motion in a homogeneous environment, when the input signal dynamics is independent of the random walker motion. We present our exact calculation for the probability distribution of signal variable, run duration distribution of the random walker and variation of mean run duration and tumble duration as a function of signaling noise in this section. In Sec. \ref{sec:px} we consider a spatially varying nutrient environment and present our numerical results for the position distribution of the random walker. A summary and few concluding remarks are presented in Sec. \ref{sec:con-ch6}.

\section{Run-and-tumble motion in a homogeneous environment} 
\label{sec:model6}

Consider a one dimensional random walker with two possible modes: run and tumble. During a run, the random walker moves with a fixed velocity along one particular direction, in this case, left or right. During a tumble, the random walker simply stays put at its current position. At the beginning of each new run, the random walker decides at random whether to run leftward or rightward. The switching between the two modes is controlled by a signal $y(t)$ whose stochastic time evolution can be written down (see below). If $y(t)$ crosses $y_0 + \Delta /2$ value from below, and the walker is in the run state, then it switches to tumble mode with probability $1$. If it is already in the tumble state, then nothing happens. Similarly, a tumbler changes to a runner with probability $1$ when $y(t)$ crosses $y_0 - \Delta /2$ from above. But at the time of crossing, if the walker is in the run mode, nothing happens.  Clearly, for $y(t) < y_0 - \Delta /2$, the random walker can only have the run mode and for $y(t) > y_0 + \Delta /2$, only tumble mode can exist. In the range $y_0 - \Delta /2 < y(t) <  y_0 + \Delta /2$, both modes can exist. Note however, that no switching event can take place in this range. When $y(t)$ enters the range through one end, and exits the range through a different end, switch happens at the time of exit. We have illustrated this process in Fig. \ref{fig:model}.

It follows from the above description that our run-and-tumble dynamics is actually different from that of an {\it E. coli} cell. Since we consider only switching events with probability one, there is no additional source of stochasticity in our run-and-tumble motion, apart from that present in the stochastic time-evolution of $y(t)$. For a given time-series of $y(t)$, it is already fixed which modes are present at what times. We will show below that this makes it possible for us to calculate many things exactly in our system. For an {\it E. coli} cell, on the other hand, switching probabilities are sharply varying, but continuous function of the CheY-P concentration \cite{cluzel2000}, and it is possible to have a switching event with small probability, which introduces another source of noise in the cell trajectory.

 In our model, we  use the same dynamics of $y(t)$ as that of CheY-P protein concentration inside an {\it E. coli} cell moving in a homogeneous nutrient background. Although the run-and-tumble motion studied by us, is not exactly same as that found in an {\it E. coli} cell, it is still interesting to see how our run-and-tumble system behaves when it receives input from the same type of a stochastic signal. In Appendix \ref{appendixA} we have derived the time-evolution equation for $y(t)$ from the biochemical pathway inside an {\it E. coli} cell and it has the form
\be
\frac{dy}{dt}=q(1+\frac{r\lambda}{2})\frac{y(1-y-wy)(1-y-2wy)}{1-y}+ry(1-y-wy)\eta(t)
\label{eq:sdea2} 
\ee
where $q,r,w$ are all constants that depend on several biochemical rate parameters, as defined in Appendix \ref{appendixA}. $\eta(t)$ is a Gaussian white noise with strength $\lambda$. To monitor the effect of input signal fluctuations on the run-and-tumble dynamics, we vary $\lambda$ in our simulations.

The value $y_0$, then naturally corresponds to that value of CheY-P concentration for which CW bias has the value $1/2$. This value turns out to be about $3.1 \mu M$ \cite{cluzel2000}. The total concentration of CheY protein in a cell is $\sim 9.7 \mu M$ \cite{b10}. Since $y(t)$ in Eq. \ref{eq:sdea2} stands for the ratio of CheY-P and CheY concentration (see Appendix \ref{appendixA}), we have $y_0 = 0.32$. Moreover, as discussed in Sec. \ref{sec:intro}, in order to ensure that the switching process is sufficiently smooth, and two switching events are separated from each other by a minimum time interval, we choose a small width $\Delta$ around $y_0$ that separates the two switching events from run to tumble, and from tumble to run. Here we present data for $\Delta =0.016$ and we have also verified (data not shown here) that our conclusions do not change for different choices of $\Delta$.

In our simulations, we consider a one dimensional box of length $L$, at the two ends of which there are reflecting boundary walls. In a time-step $dt$, the random walker in the run mode moves a distance $vdt$ where $v$ is the run speed. In a tumble mode, there is no displacement. After each tumble the random walker  will choose its direction randomly. Throughout the work we have used $L=10000 \mu m$, $v=10 \mu m /s$, $dt=0.001s$. One point about the choice of $\lambda$ range should be mentioned here. A very large $\lambda$ increases fluctuations in $y(t)$ so much that it crosses the $\Delta$ range too frequently, affecting smoothness of the underlying process. On the other hand, a very small $\lambda$ makes the $y$-distribution too narrow and $y(t)$ hardly leaves the $\Delta$ range. For our choice of $\Delta$, we find $0.001 \leq \lambda \leq  0.1$ to be suitable range. 

In the remaining part of this section, we present our exact calculations and numerical simulation results on various quantities.


\subsection{Steady state probability distribution of $y$ in run and tumble modes} \label{sec:dist}

In this section, we calculate the probability to find the cell in run and tumble modes for a given value of the stochastic signal $y$. Let $P(y,t)$ be the probability distribution of $y(t)$. From Eq. \ref{eq:sdea2} we can construct the Fokker-Planck equation for $P(y,t)$ using Ito prescription \cite{gardiner}
\begin{equation}
\frac{\partial P(y,t)}{\partial t}=-\frac{\partial}{\partial y}[B_1(y)P(y,t)]+\frac{1}{2}\frac{\partial^2}{\partial y^2}[B_2(y)P(y,t)],
\label{eq:fpe2}
\end{equation}
where, $B_1(y)=q(1+\frac{r\lambda}{2})\frac{y(1-y-wy)(1-y-2wy)}{1-y}$ and $B_2(y)=r^2k_R\lambda y^2(1-y-wy)^2 $. In steady state, left hand side of Eq. \ref{eq:fpe2} vanishes. Also, by definition, $y$ can not become negative. We show in Appendix \ref{appendixA} that $y$ actually remains bounded between $0$ and $y_m=1/(1+w)$. Therefore, we use reflecting boundary conditions at $y=0$ and $y=y_m$ which gives the following solution in steady state
\be
P(y)=\frac{w^\kappa(1-y)^{-2\kappa}[y(1-y-wy)]^{\kappa-1}}{{\cal B}(\kappa)},
\label{eq:fa2}
\ee
where ${\cal B}(\kappa)=\int^\infty_0(x(1-x))^{\kappa-1}dx=\frac{\Gamma[\kappa]^2}{\Gamma[2\kappa]}$ and $\kappa=2/(r \lambda )$. 

\begin{figure}[h!]
\includegraphics[scale=1.2]{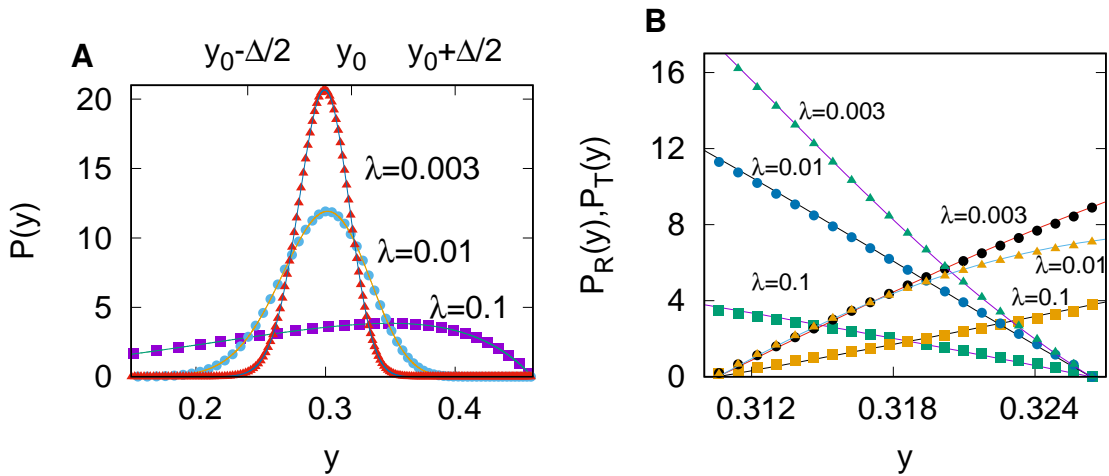}
\caption{{\bf Steady state probability distribution of the signal variable.} {\bf A:} For different noise strength $\lambda$,  probability density of   the signal variable $y$ is plotted against $y$.  Discrete points are from simulation and continuous lines are from analytical calculation using Eq. \ref{eq:fa2}. {\bf B:} Probability density $P_R(y)$ and $P_T(y)$ to observe a runner and a tumbler, respectively,  with a given $y$ value is plotted in the range $[y_0 -\Delta/2, y_0+\Delta/2]$. The decreasing curves correspond to $P_R(y)$ and increasing curves are for $P_T(y)$. The discrete points from simulations show excellent agreement with continuous lines from analytics. The probabilities are normalized, although for some $y$ values, the probability densities exceed unity.  All simulation parameters are as specified in Sec. \ref{sec:model6} and Appendix \ref{appendixA}.}
\label{fig:pa}
\end{figure}

In Fig. \ref{fig:pa}A we compare this result against numerical simulation and find good agreement for different values of the noise strength $\lambda$. In the right panel of the same figure we plot the individual probability of finding the random walker in run-state and in tumble-state for a given value of $y$, after steady state has been reached. We denote the run-state probability by $P_R(y)$ and the tumble-state probability by $P_T(y)$, and clearly, $P_R(y) + P_T(y) = P(y)$. Now, as follows from our dynamical rules, as $y$ falls below the value $y_0-\Delta /2$, tumble modes can not exist and the random walker is always in the run mode, {\sl i.e.} $P_R(y) = P(y)$ for $y \leq y_0-\Delta /2$. Similarly, for $y \geq y_0+\Delta /2$, we have $P_T(y) = P(y)$ and $P_R(y) =0$. Both $P_R(y)$ and $P_T(y)$ have non-zero values for $y_0 - \Delta /2 < y < y_0 + \Delta /2$. To solve for $P_R(y)$ in this range, we notice that it follows the same Fokker-Planck equation as Eq. \ref{eq:fpe2} and in steady state this equation has the general solution
\begin{equation}
P_R(y)=\frac{w}{(1-y)^2}\left [ \frac{1-(\frac{2wy}{1-y}-1)^2}{4} \right ]^{\kappa/2-1} \left [ C_1P^\kappa_\kappa \left ( \frac{2wy}{1-y}-1 \right )+C_2 Q_\kappa ^\kappa \left ( \frac{2wy}{1-y}-1 \right ) \right ],
\label{eq:pra}
\end{equation}
where $P^\kappa_\kappa $ and $Q_\kappa^\kappa$ are  associated Legendre polynomial of first and second kind, respectively. The constants $C_1$ and $C_2$ can be determined from the boundary conditions $P_R(y_0-\Delta /2)=P(y_0-\Delta /2)$ and $
P_R(y_0+\Delta /2)=0$, discussed above. $P_T(y)$ can simply be obtained from $P_T(y) = P(y) - P_R(y)$. In Fig. \ref{fig:pa} we verify our analytical calculation against numerical simulations for few different values of the noise strength $\lambda$ and find good agreement. 


\subsection{Average run and tumble duration decreases with signaling noise} 
\label{sec:avg-run}

\begin{figure} [h!]
\includegraphics[scale=1.2]{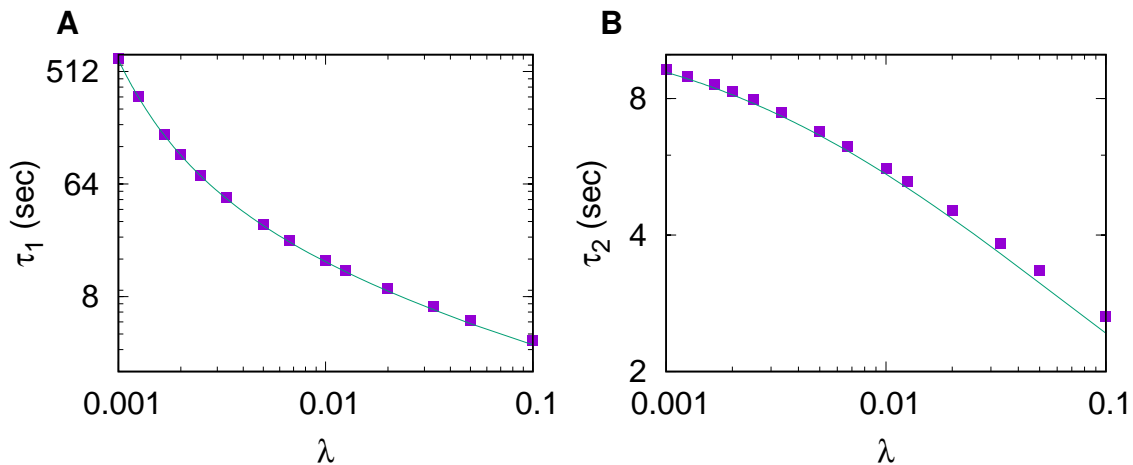}
\caption{{\bf Average run and tumble duration as a function of signaling noise strength $\lambda$.} {\bf A.} The average run duration $\tau_1$ decreases as a function of $\lambda$. The range of variation of $\tau_1$ is quite significant. {\bf B.} The average tumble duration $\tau_2$ decreases  with $\lambda$ but the range of variation is much smaller than that for $\tau_1$. Discrete points are from simulations and continuous lines are from analytics. The  simulation parameters are same as in Fig. \ref{fig:pa}.}
\label{fig:avgtau1}
\end{figure}

One possible way to characterize a run-tumble motion is by measuring the average duration of a run mode and a tumble mode. In Fig. \ref{fig:avgtau1}A we plot average run duration as a function of the noise strength $\lambda$. We find that as signaling noise decreases, the average run duration increases. In fact for low $\lambda$ values, average run duration becomes so large that in our simulations we have to consider large system size $L$ to avoid finite size effects. Fig. \ref{fig:avgtau1}B shows variation of average tumble duration with noise. Below we discuss how to calculate these averages exactly. 

Note that at the beginning of a run, i.e. just at the instant when tumble to run switch happens, the input signal $y$ always has the value $y_0 - \Delta /2$. Starting from this value, when $y$ crosses $y_0 + \Delta /2$ for the first time, the run ends and a tumble begins.  Therefore, a run can be viewed as a first passage event in the $y$-space. This makes it possible to calculate the average run duration and even the run-length distribution (see next subsection) exactly. If $T(y_i, y_f)$ denotes the mean first passage time for $y$ to reach the value $y_f$ for the first time, starting from an initial value $y_i$, then $T(y_0 - \Delta/2, y_0 + \Delta/2)$ represents the mean run duration and $T(y_0 + \Delta/2, y_0 - \Delta/2)$ stands for the mean tumble duration.

Let $p(y',t|y,0)$ be the conditional probability that the input signal has the value $y'$ at time $t$, given that it started with the value $y$ at time $t=0$. This conditional probability follows the backward Fokker-Planck equation \cite{gardiner,risken}
\begin{equation}
\frac{\partial p(y',t|y,0)}{\partial t}=B_1(y) \frac{\partial p(y',t|y,0)}{\partial y} + \frac{1}{2}B_2(y) \frac{\partial^2 p(y',t|y,0)}{\partial y^2}
\label{eq:fpe4}
\end{equation}
where $B_1(y)$ and $B_2(y)$ are drift and diffusion terms appearing in Eq. \ref{eq:fpe2}. To calculate the mean first passage time at $y_0+\Delta /2$, starting from $y_0-\Delta /2$, we put an absorbing boundary condition at the target $y=y_0+\Delta /2$ and remember the reflecting boundary condition at $y=0$. The survival probability $G(y,t; y_0+\Delta /2)$ is defined as the probability that starting from $y<y_0+\Delta /2$, the signal variable has not reached the target value $y_0+\Delta /2$ till time $t$. Clearly, $G(y,t;y_0+\Delta /2) = \int_0 ^ {y_0+\Delta /2} dy' p(y',t|y,0)$. From Eq. \ref{eq:fpe4} it follows that $G(y,t; y_0+\Delta /2)$ satisfies the following equation
\be 
\frac{\partial G(y,t; y_0+\Delta /2)}{\partial t} = B_1(y) \frac{\partial G(y,t; y_0+\Delta /2)}{\partial y} + \frac{1}{2}B_2(y)\frac{\partial^2 G(y,t; y_0+\Delta /2)}{\partial y^2} \label{eq:G}
\ee
with the initial condition $G(y,0; y_0+\Delta /2) =1$ and the reflecting and absorbing boundary conditions are implemented as $\partial_y G(y,t; y_0+\Delta /2)|_{y=0} =0$ and $G(y,0; y_0+\Delta /2)|_{y=y_0+\Delta /2} =0$. The survival probability  till time $t$ can be alternatively stated as the probability that the first passage time is larger than $t$. Therefore, the first passage time distribution is simply $-\partial_t G(y,t; y_0+\Delta /2)$. The mean first passage time is then $T(y,y_0+\Delta /2) = -\int_0 ^{\infty} dt \; t \;\partial_t G(y,t; y_0+\Delta /2) = \int_0^\infty dt G(y,t; y_0+\Delta /2) $ which follows the equation
\be 
B_1(y) \frac{\partial T(y,y_0+\Delta /2)}{\partial y} + \frac{1}{2}B_2(y)\frac{\partial^2 T(y,y_0+\Delta /2)}{\partial y^2} = -1.
\ee
This equation can be solved to get the mean run duration as 
\be
T(y_0-\Delta /2, y_0+\Delta /2) =\tau_1= 2\int_{y_0-\Delta /2} ^ {y_0+\Delta /2} \frac{dy}{\psi(y)} \int_0 ^ y \frac{\psi(z)}{B_2(z)},
\label{eq:t1}
\ee 
where $\psi(x) = \exp \left [ \int_0 ^ x dx' 2B_1(x') / B_2(x') \right ] $. 
Similarly, mean tumble duration can be written as 
\be
T(y_0+\Delta /2, y_0-\Delta/2) =\tau_2= 2\int_{y_0-\Delta /2} ^ {y_0+\Delta /2} \frac{dy}{\psi(y)} \int_y ^ {y_{m}} \frac{\psi(z)}{B_2(z)}.
\label{eq:t2}
\ee 
where reflecting boundary condition is used for $y=y_m$ and absorbing boundary condition for $y=y_0-\Delta/2$. We find good agreement with the simulation data in Fig. \ref{fig:avgtau1}.


\subsection{Distribution of the run duration of the random walker}
\label{sec:dist-run}

\begin{figure}[h!]
\includegraphics[scale=1.3]{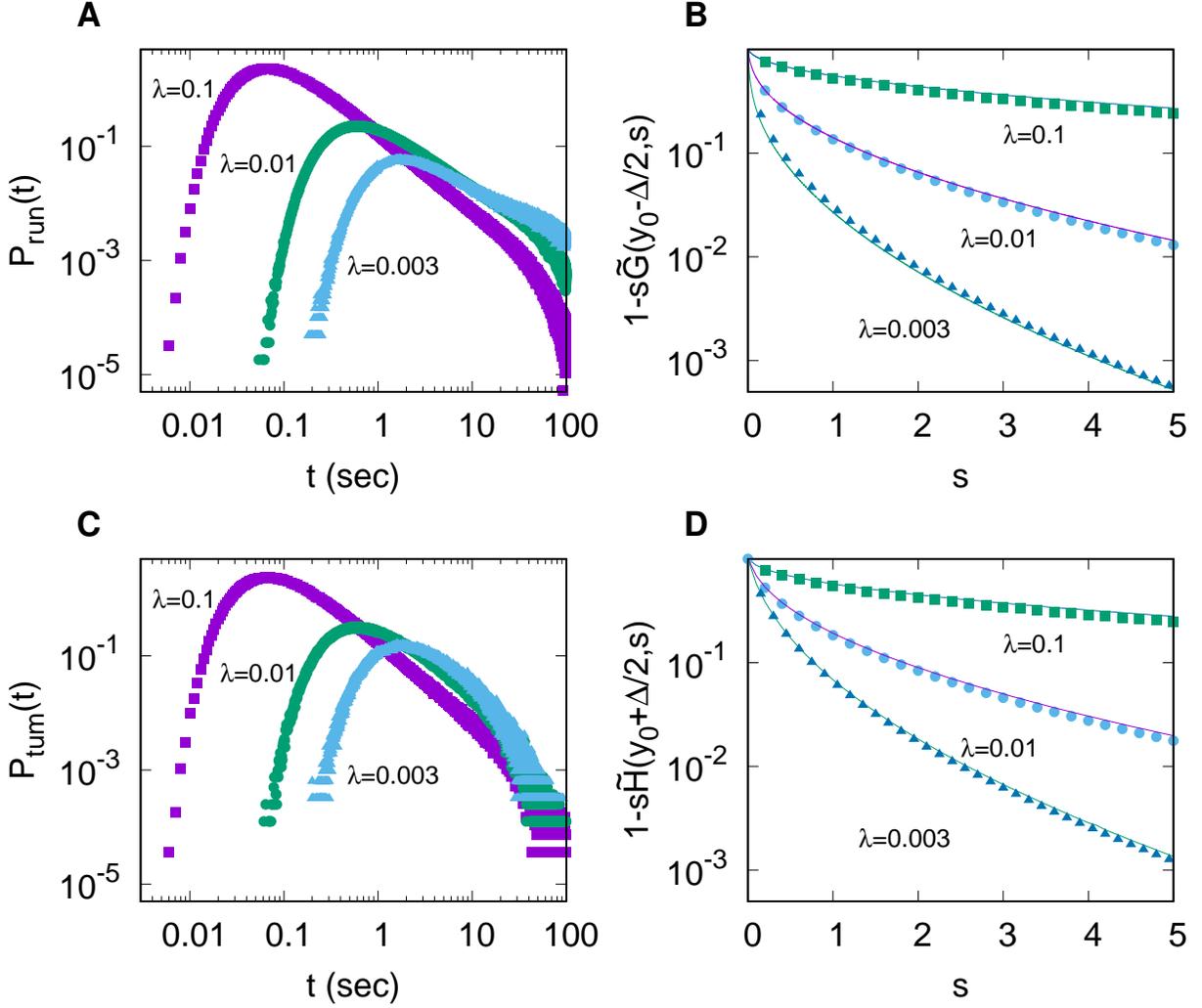}
\caption{{\bf Distribution of run duration and tumble duration for different value of noise strength.} {\bf A:} Simulation results for the distribution of the run duration of the bacterium $P_{run}(t)$. The distribution has a peak whose position shifts leftward as noise increases. {\bf B:}
The Laplace transform of $P_{run}(t)$ analytically calculated and plotted in continuous lines. The discrete points show Laplace transform calculated from the data in panel A and we find good agreement.{\bf C:} Simulation results for the distribution of the tumble duration of the bacterium $P_{tum}(t)$. {\bf D:}
The Laplace transform of $P_{tum}(t)$ analytically calculated and plotted in continuous lines and the discrete points are the  Laplace transform of data in panel C. The simulation parameters are same as in Fig. \ref{fig:pa}}
\label{fig:distau}
\end{figure}

Using the correspondence between the run and tumble durations of the random walker and the first passage events for the input signal, it is possible to calculate not only the average run and tumble durations, but also the full distribution function of these durations. We outline this calculation in this subsection. First we present our numerical data for the run duration distribution. In Fig. \ref{fig:distau}A we plot the probability $P_{run}(t)$ that the random walker has a residence time $t$ in the run mode, for different values of the  noise strength $\lambda$. We find that the probability vanishes for very small and large $t$, and shows a peak in between. The peak position depends on $\lambda$ and as $\lambda$ increases, the peak shifts towards smaller values of $t$. In other words, the most probable run duration becomes smaller and smaller as noise increases. This behavior is similar to that of the mean run duration shown in Fig. \ref{fig:avgtau1}. As noise increases, the signal $y(t)$ takes less and less time to reach the value $y_0 + \Delta /2$, starting from  $y_0 - \Delta /2$ since the diffusivity $B_2(y)$ becomes larger with noise. 

To calculate the run duration distribution analytically, we focus on its Laplace transform. First we consider the Laplace transform of the survival probability 
$\tilde{G}(y,s)=\int_0^\infty dt \; e^{-st}\; G(y,t; y_0+\Delta/2)$, where for simplicity of notation we have dropped $y_0+\Delta/2$ from the argument of the $\tilde{G}$. From Eq. \ref{eq:G} it follows that 
\begin{equation}
 \frac{1}{2}B_2(y) \partial_y^2 \tilde{G}(y,s)+B_1(y) \partial_y \tilde{G}(y,s)-s\tilde{G}(y,s)=-1,
 \label{eq:lapG}
\end{equation}
Defining $\tilde{U}(y,s)=\tilde{G}(y,s)-\frac{1}{s}$ we get
\begin{equation}
\frac{1}{2}B_2(y)\partial_y^2 \tilde{U}(y,s)+B_1(y) \partial_y \tilde{U}(y,s)-s\tilde{U}(y,s)=0.
\label{eq:lapU}
\end{equation}
whose general solution is
\begin{equation}
\tilde{U}(y,s)=\left [ \frac{wy(1-y-wy)}{(1-y)^2} \right ]^{\kappa/2} \left [ D_1P^{\sqrt{\kappa^2+4\mu(s)}}_\kappa \left ( \frac{2wy}{1-y}-1 \right )+D_2 Q_\kappa ^{\sqrt{\kappa^2+4\mu(s)}} \left ( \frac{2wy}{1-y}-1 \right ) \right ],
\end{equation}
where $\mu(s)=\frac{2s}{\lambda qr}$. The constants $D_1$ and $D_2$ can be determined from the boundary conditions: $\tilde{G}(y_0+\Delta /2,s)=0$ and $\partial_y\tilde{G}(y,s)|_{y=0}=0$ for all $s$. The Laplace transform of first passage time distribution is given by $1-s\tilde{G}(y,s)$ which can be evaluated at $y = y_0 - \Delta/2 $ to obtain the Laplace transform of run-length distribution. We compare our calculation with simulation results in Fig. \ref{fig:distau}B and find good agreement. In Figs. \ref{fig:distau}C and \ref{fig:distau}D we similarly plot distribution of tumble duration and its Laplace transform. Note that the main difference between $P_{run}(t)$ and $P_{tum}(t)$ can be seen for large $t$ values, when $P_{tum}(t)$ decays more sharply. As a result mean run-duration $\tau_1$ is always larger than mean tumble duration $\tau_2$ (see also the data in Figs. \ref{fig:avgtau1}A and \ref{fig:avgtau1}B).

\section{Run and Tumble motion in an environment with spatial variation}
\label{sec:px}

In the previous section, we studied the situation, when the coupling between the stochastic signal $y(t)$ and the random walk motion is one way. While the random walk switches between the run and tumble modes depending on the value of the signal, the signal itself fluctuates independently according to Eq. \ref{eq:sdea2}. In this section, we consider a two-way coupling between the signal dynamics and the random walker motion. More specifically, we consider a time-evolution equation for $y(t)$ which involves the position $x$ of the random walker as well. Thus, the random walker runs and tumbles following the $y(t)$ value as before, but the random walker position now influences the time-evolution of $y(t)$. We write the equation for $y(t)$ dynamics as
\be
\frac{dy}{dt}=q(1+\frac{r\lambda}{2})\frac{y(1-y-wy)(1-y-2wy)}{1-y}-s\frac{y(1-y-wy) }{(K_A+c(x))(K_I+c(x))}+ry(1-y-wy)\eta(t).
\label{eq:dydt-position} 
\ee
As explained in Appendix \ref{appendixA}, this dynamics is borrowed from a physical system that describes the fluctuation in the CheY-P protein level inside an {\it E. coli} cell in presence of a nutrient concentration gradient in the extra-cellular environment. For Eq. \ref{eq:dydt-position} we have chosen a nutrient concentration profile that is linear and has the form $c(x) = c_0 (1+x/x_0)$. The run-and-tumble motion of an actual {\it E. coli} cell in such a nutrient environment gives rise to chemotaxis and in the long time limit there is larger probability to find the cell at regions with higher $c(x)$ \cite{degennes, jiang, sc, sdev}.

\begin{figure}
\includegraphics[scale=1.4]{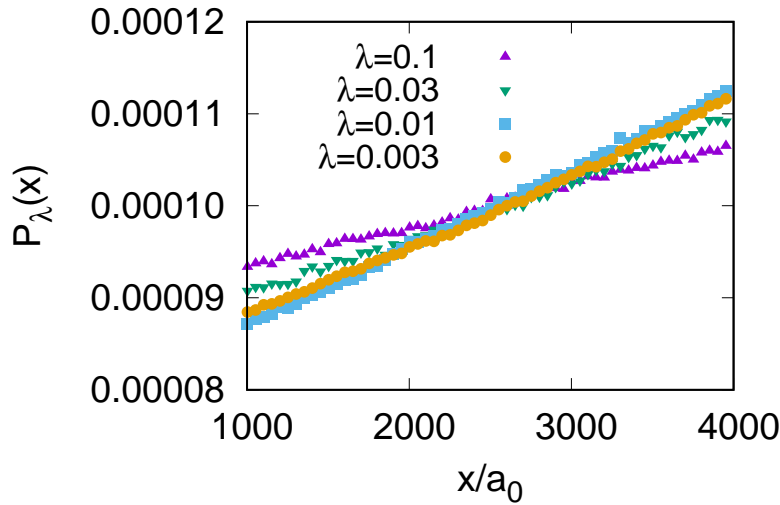}
\caption{{\bf The distribution $P_\lambda(x)$ of the random walker position $x$ for different noise strengths.}  The x-axis has been scaled with the size of the cell which is $a_0=2 \mu m$. We have used $c(x)=c_0(1+x/x_0)$ here and for all $\lambda$ values, $P_\lambda(x)$ shows a positive slope. For large $\lambda$, the slope is less. We have chosen $c_0=200 \mu M$ and $x_0=200000 \mu m$ here and all other parameters are as in Fig. \ref{fig:pa}.}
\label{fig:px}
\end{figure}

The two-way coupling between $y(t)$ and $x(t)$ makes it difficult to obtain analytical solution in this case and we study the system using numerical simulations. We consider the weak gradient limit here when $c(x)$ varies very slowly with $x$. As a result, the quantities like $P_{run}(t)$ or $P_{tum}(t)$ look almost similar to our data in Fig. \ref{fig:distau}. We do not present these data here. However, it is interesting to see whether in the long time limit the random walker manages to localize itself in the region with large $c(x)$. In Fig. \ref{fig:px} we show the data for the position distribution of the random walker in the long time limit. We find that $P_{\lambda}(x)$ increases with $x$, roughly linearly. This result shows that although the run-and-tumble dynamics is significantly different from and simpler than that of an {\it E. coli} cell, the walker still manages to locate itself in the region with higher nutrient concentration with larger probability. Our data show that $P_{\lambda}(x)$ varies as $c(x)$ for small and intermediate $\lambda$ values. However, when $\lambda$ becomes large, $P_{\lambda}(x)$ gradually becomes flat, as expected in the limit of large signaling noise, when the time-evolution of $y(t)$ is mainly governed by the stochastic fluctuations, and its $x$-dependence can be almost ignored. 

\section{Conclusion}
\label{sec:con-ch6}

In this paper, we have investigated the effect of a sharp step-like response function on a run-and-tumble random walk. In nature run-and-tumble motion is ubiquitous in a wide variety of organisms. While an intra-cellular biochemical reaction network controls the motion in all these cases, some organisms, for example, {\it E. coli} bacteria, show a particularly sensitive dependence on these reactions. The transition rate of an {\it E. coli} cell from run mode to tumble mode depends strongly and sensitively on the fluctuating concentration of the motor protein CheY-P, which is an important component of its reaction network. This motivates a general theoretical question that we consider in this paper: what happens when a run-and-tumble motion is coupled to a stochastic input signal via a sensitive response. We are interested in two different cases: one in which the stochastic dynamics of the input signal is an independent process and another in which the signal variable dynamics also depends on the spatial location of the random walker. In the first case, we specifically choose the signal variable dynamics from that of CheY-P protein concentration for an {\it E. coli} cell in a homogeneous nutrient environment. The simple switching dynamics that we use for our run-and-tumble walker makes it possible to calculate many things exactly in this case. In the second case, we consider a signal variable whose time-evolution  mimics CheY-P dynamics for an {\it E. coli} cell in a spatially varying nutrient environment. Interestingly, our numerical simulations show that even with its simple run-and-tumble strategy, the random walker manages to localize in a region where nutrient density is higher.

The run-and-tumble motion that we consider here, is significantly different from that executed by an {\it E. coli} cell. While for an {\it E. coli} cell, the tumbling bias varies sensitively, but continuously as a function of the CheY-P level, in our model the switching probability between the run and tumble modes show a sharp jump from $0$ to $1$. This allows us to address the theoretical question of the effect of sharp response in the simplest possible setting. Although our results in Fig.  \ref{fig:px} show that the basic signature of chemotaxis is still retained in our model, we also find some important differences from well-known {\it E. coli} behavior. One such crucial difference is observed in run duration distribution. For low signaling noise, the run and tumble duration follow Poisson process , so that {\it E. coli} shows exponential distribution of run and tumble duration. As the signaling noise gets larger, longer runs become more probable and the distribution of run changes to a power law but tumble duration still shows exponential distribution \cite{korobkova, tu2005, matthaus2009, matthaus2011, vasily, dev18}. In contrast, in our model, runs and tumbles can be described as first passage events and both the distribution of run and tumble (see Fig. \ref{fig:distau}A and C) have a peak and shows a power law tail for all values of $\lambda$, indicating that this process never follows Poission process. Moreover, we also find that with increasing noise, longer runs become less probable in our case. As noise level becomes lower, the mean run duration in our model increases rather strongly. For {\it E. coli} motion also mean run duration becomes larger for lower signaling noise, but the variation is much weaker in that case \cite{dev18}.

As we mentioned in Sec. \ref{sec:model6}, the run-and-tumble trajectory of {\it E. coli} has an additional level of stochasticity coming from the fact that switching probability can be less than one, which means for a given time-series of the input signal, it is possible to generate different run-tumble trajectories. However, in our model, switching probability is either $0$ or $1$ and can be nothing in between. This deterministic nature means that only one run-and-tumble trajectory is possible for a given signal time-series. Although the direction of a new run is still chosen randomly at the time of every tumble to run switch in our model, but in a homogeneous nutrient background it makes no difference whether the random walker is running towards left or right. The differences mentioned in the previous paragraph may be alternatively viewed as the result of this deterministic vs stochastic aspect. It also shows that although CW bias of {\it E. coli} increases really sharply as CheY-P level changes, when that response is actually replaced by a jump in the switching probability, system shows qualitatively different behavior in many aspects. It may be interesting to gradually vary the steepness of a sigmoidal response curve and see if there is a crossover between the two behaviors.

Our model of run-and-tumble motion complements the widely used  coarse-grained model where instead of looking at the switching events between the run and tumble modes, the system is described over a time-scale in which a large number of switching events have already taken place. This coarse-graining allows one to describe the motion in terms of standard drift-diffusion process \cite{cates,sc,tailleur}. Using this formalism, the motion of {\it E. coli} in a homogeneous nutrient environment, can be described as an unbiased diffusion. Interestingly, the diffusion co-efficient in this case is order of magnitude larger than that expected for an ordinary Brownian motion of a particle whose size is comparable to that of a bacterial cell \cite{cates}. Contrary to this coarse-grained approach, our model probes a run-and-tumble dynamics at a more microscopic level, where each switching event is taken into account and the interval between two successive switching events is described using time-evolution of a stochastic signal. Of course, in the very long time limit, even our model yields diffusive behavior (data not shown here) for the case of homogeneous environment discussed in Sec. \ref{sec:model6}. For the spatially varying environment discussed in Sec. \ref{sec:px}, the random walker picks up a drift velocity which is proportional to the spatial gradient of $c(x)$, and this is consistent with earlier known results \cite{degennes,sc}.

Finally, at a more general level, many different organisms, other than  {\it E. coli}, show run-and-tumble motility. Some of these organisms have very similar motility mechanism as {\it E. coli}, {\sl e.g.} {\it Salmonela typhimurium} \cite{salmonela}, {\it Bacillus subtilis} \cite{bacilus} or {\it Serratia marcescens} \cite{serra}, but prokaryotes like  {\it Rhodobacter sphaeroides} have a somewhat different mechanism. In a {\it Rhodobacter sphaeroides} cell, a single flagellum is present and CCW rotation of the motor causes a run, while abrupt ceasing of rotation allows the cell to tumble or reorient \cite{rhodo}. Among eukaryotic cells, {\it Chlamydomonas rheinhartii} contains two flagella and when these two flagella beat synchronously, the cell swims smoothly, while an asynchronous beating results in tumbles \cite{chlamy}; {\it Tritrichomonas foetus} has four flagella and they follow two distinctly different beating patterns in order to cause run mode and tumble mode of the cell motion \cite{tritrich}. Many of these organisms are experimentally not as well-characterized as {\it E. coli}. But in all cases the switching between the run mode and tumble mode, are controlled by flagellar motion, which in turn depends on intra-cellular signaling. Therefore, a general understanding of how a sensitive dependence on the stochastic signal affects a run-and-tumble motion may prove useful for these systems as well and our study takes a step in this relatively less-explored direction.

\section*{Acknowledgements} 
SC acknowledges financial support from the Science and Engineering Research Board, India (Grant No. EMR/2016/001663). The computational facility used in this work was provided through the Thematic Unit of Excellence on Computational Materials Science, funded by Nanomission, Department of Science and Technology (India).

\appendix

\section{The signaling pathway inside an {\it E. coli} cell}
\label{appendixA}
\renewcommand{\theequation}{A-\arabic{equation}}
\renewcommand{\thefigure}{A-\arabic{figure}}
\setcounter{equation}{0}
\setcounter{figure}{0}

The signaling pathway inside an {\it E. coli} cell can be described in terms of three coupled dynamical variables: the activity $a(t)$ and methylation level $m(t)$ of the chemo-receptor complex, CheY-P concentration $y(t)$. We use the standard model introduced in \cite{jiang,tu2008} and subsequently modified in \cite{flores}.

The activity is defined as the probability to find the chemo-receptor in the active state and has the expression
\begin{equation}
a=\dfrac{1}{1+e^{N (f_m + f_c)} }, \label{eq:act}
\end{equation} 
with $f_m=\alpha(m_0-m)$ and $f_{c}=-\log \left (\dfrac{1+c(x)/K_A}{1+c(x)/K_I} \right )$ \cite{tu05,wingreen06}. Here, $c(x)$ is the concentration of the nutrient at the cell position $x$. Clearly, as the cell position $x$ or methylation level $m$ change with time, activity $a$ also changes. However, by definition, $a(t)$ always stays bounded between $0$ and $1$. The parameter values are $N=6$, $K_A= 3 mM $, $K_I = 18.2 \mu M$, $\alpha =1.7$, $m_0=1$ \cite{jiang,flores}.

The (de)methylation reaction is the slowest reaction step in the biochemical pathway. Hence any stochastic fluctuation that happens at this step, is propagated downstream as a slow noise and cannot be integrated out. Because of this, the signaling noise is often incorporated as an additive Gaussian white noise in the methylation reaction \cite{flores}
\begin{equation}
\dfrac{dm}{dt}=k_R(1-a)-k_Ba+\eta(t). \label{eq:meth}
\end{equation}
Here, $\eta (t)$ denotes stochastic noise with properties $<\eta>=0$ and $<\eta(t)\eta(t^\prime)>= \lambda (k_R(1-\bar{a})+k_B\bar{a})\delta(t-t^\prime)$, where $\bar{a}=1/2$ is the average activity level in absence of any noise. $k_R$ and $k_B$ are the rate parameters of the reactions in the biochemical pathway. They have small values, which makes the above reaction a slow one. In our simulation, we have used $k_R=k_B=0.015 s^{-1}$ \cite{flores,shimizu2010}, which gives $<\eta(t)\eta(t^\prime)>= \lambda k_R \delta(t-t^\prime)$.

Fluctuations in methylation level will also cause fluctuations in activity which in turn affects the phosphorylation of CheY proteins. In the phosphorylated state, CheY-P proteins bind to the flagellar motors and cause the cell to tumble. Denoting the fraction of phosphorylated CheY proteins as $y$, we can write \cite{flores}   
 \begin{equation}
\dfrac{dy}{dt}=k_Ya(1-y)-k_Zy \label{eq:yeq}
\end{equation}   
where the rates $k_Y=1.7 s^{-1}$ and $k_Z=2 s^{-1}$ are much larger than the
(de)methylation rates \cite{tu2008,flores}.

In the case when the cell moves in a homogeneous nutrient background, $c(x)=c_0$, the activity $a(t)$ becomes a function of $m(t)$ alone, and from Eqs. \ref{eq:act} and \ref{eq:meth} and 
one can write
\begin{equation}
\frac{da}{dt}=k_RN\alpha a(1-a)(1-2a)(1+\frac{N\alpha\lambda}{2})+N\alpha a(1-a)\eta(t)
\label{eq:sde_a} 
\end{equation}
A quasi steady state approximation can be made at this stage, using the fact that the $y$-dynamics is sufficiently fast, and hence at the time-scale over which $a(t)$ is changing, an average $y$ concentration is felt, which gives $y(t)=a(t)/(a(t)+k_Z/k_Y)$. Then Eq. \ref{eq:sde_a} becomes
\be 
\frac{dy}{dt}=k_RN\alpha(1+\frac{N\alpha\lambda}{2})\frac{y(1-y-\frac{k_Zy}{k_Y})(1-y-\frac{2k_Zy}{k_Y})}{1-y}+N\alpha y(1-y-\frac{k_Zy}{k_Y})\eta(t).
\label{eq:sde_y} 
\ee
Writing $q=k_RN\alpha$, $r=N\alpha$ and $w=k_Z/k_Y$, we get Eq. \ref{eq:sdea2}. In the case when the nutrient concentration is not uniform, but varies linearly in space, $c(x) = c_0 (1+x/x_0)$, activity $a(t)$ changes when the methylation level changes, or when the cell moves in the medium. In that case, Eq. \ref{eq:sde_y} becomes
\be
\frac{dy}{dt}=q(1+\frac{r\lambda}{2})\frac{y(1-y-wy)(1-y-2wy)}{1-y}-\frac{vNc_0}{x_0}\frac{K_A-K_I}{(K_A+c(x))(K_I+c(x))}y(1-y-wy)+ry(1-y-wy)\eta(t).
\ee
Writing $s=\frac{vNc_0}{x_0}(K_A-K_I) $ gives us Eq. \ref{eq:dydt-position}. 

Note that the quasi steady state approximation used above, means that since $a(t)$ always stays within the range $[0,1]$, the variable $y(t)$ should also stay in $[0,y_m]$, where $y_m = 1/(1+k_Z/k_Y)$. In the main paper, we provide exact solution of Eq. \ref{eq:sde_y} where we use reflecting boundary conditions at $y=0$ and $y=y_m$.


\end{document}